\documentclass[%
 letterpaper,
 amsmath,amssymb,
 aps,prl,reprint,superscriptaddress, showpacs
]{revtex4-1}
\usepackage{graphicx}
\usepackage{dcolumn}
\usepackage{bm}

\newcommand{\ket}[1]{\ensuremath{|{#1}\rangle}}

\begin{document}

\title{Photon propagation in a discrete fiber network: An interplay of coherence and losses}

\author{Alois Regensburger}
\affiliation{Institute of Optics, Information and Photonics, University Erlangen-Nuremberg, 91058 Erlangen, Germany}
\affiliation{Max Planck Institute for the Science of Light, 91058 Erlangen, Germany}

\author{Christoph Bersch}
\affiliation{Institute of Optics, Information and Photonics, University Erlangen-Nuremberg, 91058 Erlangen, Germany}
\affiliation{Max Planck Institute for the Science of Light, 91058 Erlangen, Germany}

\author{Benjamin Hinrichs}
\affiliation{Institute of Optics, Information and Photonics, University Erlangen-Nuremberg, 91058 Erlangen, Germany}
\affiliation{Max Planck Institute for the Science of Light, 91058 Erlangen, Germany}

\author{Georgy Onishchukov}
\affiliation{Max Planck Institute for the Science of Light, 91058 Erlangen, Germany}

\author{Andreas Schreiber}
\affiliation{Max Planck Institute for the Science of Light, 91058 Erlangen, Germany}

\author{Christine Silberhorn}
\affiliation{Max Planck Institute for the Science of Light, 91058 Erlangen, Germany}
\affiliation{University of Paderborn, Applied Physics, 33098 Paderborn, Germany}

\author{Ulf Peschel}%
\email{ulf.peschel@mpl.mpg.de}%
\affiliation{Institute of Optics, Information and Photonics, University Erlangen-Nuremberg, 91058 Erlangen, Germany}%

\date{\today}

\begin{abstract}
We study light propagation in a photonic system that shows stepwise evolution in a discretized environment. It resembles a discrete-time version of photonic waveguide arrays or quantum walks. By introducing controlled photon losses to our experimental setup, we observe unexpected effects like sub-exponential energy decay and formation of complex fractal patterns. This demonstrates that the interplay of linear losses, discreteness and energy gradients leads to genuinely new coherent phenomena in classical and quantum optical experiments. Moreover, the influence of decoherence is investigated.
\end{abstract}

\maketitle

Although present in almost every experiment, the role of photon losses
in optics is quite elusive. In linear wave optics, losses are often
ignored, as in many cases they can be scaled out. Mostly, losses are
considered to be a nuisance which do neither offer any benefits nor
introduce genuinely new coherent dynamics. In this Letter, we show experimentally that controlled losses generate new effects of linear wave evolution in high dimensional discrete networks. 

We demonstrate how coherence and discreteness drastically impact energy decay in such systems, reducing the losses to the sub-exponential regime. Furthermore, the interplay between coherence, discreteness and losses leads to a completely different intensity distribution over the output ports of the network. The resulting formation of triangular patterns has similarities to the Sierpinski sieve. These experimental results demonstrate that the discreteness of both evolution and transverse direction has an unexpected influence on lossy photon propagation. This gives an entirely new flavor to the role of linear losses both in classical and quantum optics. 

Our experiment can be seen as a classical realization \cite{Knight2003,Bouwmeester1999, Rohde2011a} of a quantum walk (QW) \cite{Kempe2003, Aharonov1993, Schreiber2010, Karski2009, Schmitz2009, Broome2010, Zahringer2010} of single particles and resembles a discrete time version of photonic waveguide arrays (WGA) \cite{Christodoulides2003}. Recently, WGAs have been utilized to investigate quantum propagation
\cite{Bromberg2009a,Peruzzo2010}. In quantum optics, many important
phenomena are governed by interference effects, like the Grover search
algorithm \cite{Kwiat2000a} or the QW of single particles. 
However, common WGAs do not allow for a selective management of phases
and losses during propagation. The transfer of the concept of discreteness to the temporal domain as well as an additional discretization of the propagation direction in our recirculating fiber-loop setup enables this kind of manipulation during each propagation step. This allows to explore completely new types of coherent dynamics.

\begin{figure}
\includegraphics{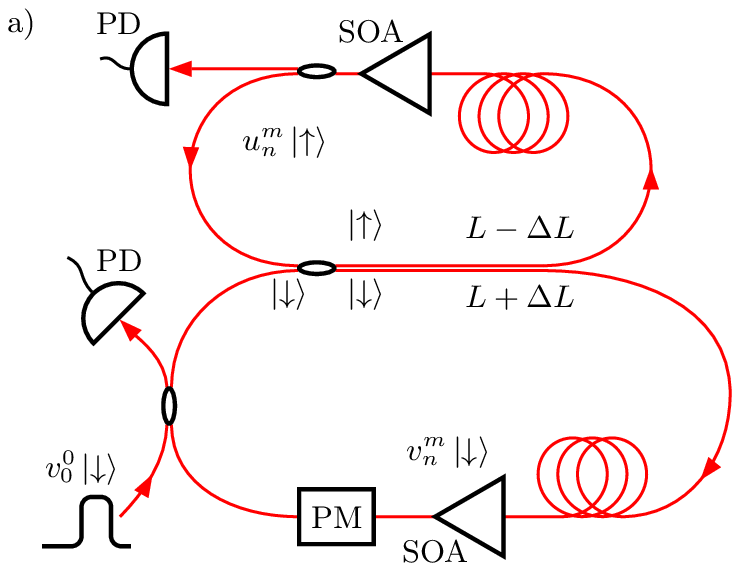}
\vspace*{5mm}
\includegraphics{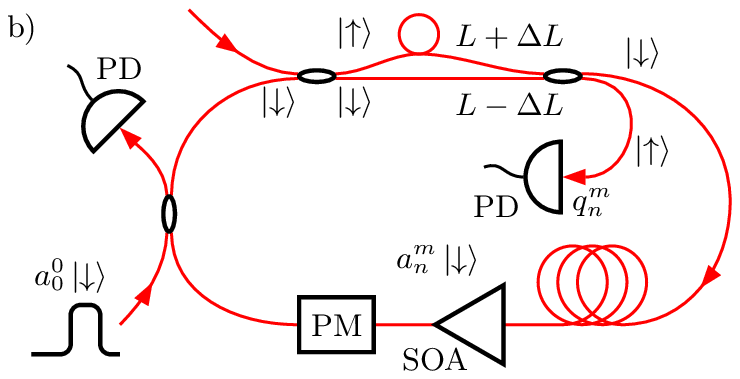}
\caption{\label{fig:loops} Principal scheme of all-fiber experimental setup (a) without deliberately introduced losses (b) to introduce a controlled loss of photons by measuring all intensity in the $\ket{\uparrow}$ state after every second 50/50 coupler; SOA: semiconductor optical amplifier; PM: phase modulator; PD: photodiode}
\end{figure}
\begin{figure*}
\includegraphics{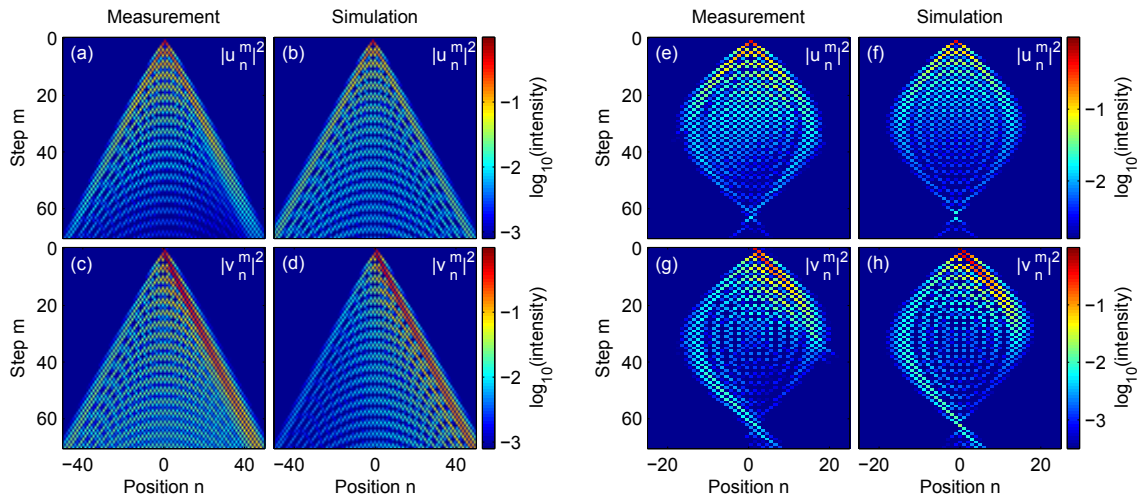}
\caption{\label{fig:qwbloch} Evolution of wave starting in state $\ket{\downarrow}$ at position $n=0$ in the lossless case (Fig. 1a); (a-d) measurements and simulations of ballistic spreading of intensities $\left| u_n^m \right|^2$ in the upper and $\left| v_n^m \right|^2$ in the lower loop without phase modulation; dynamics are the same as in a discrete time QW (e-h) discrete time Bloch oscillations for linear phase gradient $\alpha = \frac{2 \pi}{32}$ }
\end{figure*}
Moreover, the transport of excitations is largely affected by the environment in the form of decoherence, losses and energy gradients. In photosynthesis as a prominent example, the initially coherent transport of an exciton that is driven by an energy gradient even gets accelerated by the noisy environment \cite{Plenio2008a,Mohseni2008,Caruso2009}. Although surely present, the influence of particle decay has not yet been investigated so intensively. It is therefore desirable to gain a better understanding of the interplay of losses and spatially varying potentials in the dynamics of coherent processes.

In the following, we study experimentally how discreteness together with photon losses and energy gradients affect the evolution of light governed by wave interference. 

In our experimental setups (Figs. 1a,b, see \cite{suppl2} for details), the discrete spatial dimension (position space) is emulated by time-multiplexing \cite{Schreiber2010,Schreiber2011a}. In the first case (Fig. 1a), light propagates in two loops of single-mode fiber with a length difference $2 \Delta L$ which are connected by a 50/50 coupler. Each roundtrip corresponds to a discrete propagation step $m$. A coherent light pulse traveling in the longer (shorter) loop makes a step to the right (left) in position space $n$ and at the central coupler, multi-path interference takes place. The gain of each semiconductor optical amplifier in the loops is adjusted to compensate only the signal losses caused by light absorption and monitoring. The amplifiers do not affect classical wave interference and the system behaves as if it was lossless (see \cite{suppl2}). However, they allow us to realize a considerably larger field spreading than previously reported for QWs \cite{Schreiber2011a} or in WGAs, with a big potential for even a further increase of propagation steps. Certainly, their noise would affect the evolution of non-classical light states in the system.

The distribution of the signal between the upper and lower loop is represented by two states $\ket{\uparrow}$ and $\ket{\downarrow}$. At step $m=0$, a pulse is fed into the lower loop (state $\ket{\downarrow}$). The field distribution after $m$ steps is given by the amplitudes $u_n^m$ in the upper and $v_n^m$ in the lower loop for a respective position $n$. Fast photodiodes measure the intensities $\left| u_n^m \right|^2$ and $\left| v_n^m \right|^2$  probed by tap couplers. Moreover, a phase modulator is inserted into the lower loop to induce a phase or energy gradient in position space. 

The evolution of the amplitudes can be described by the algebraic equations
\begin{equation}
\begin{split}
u_n^{m+1} &= \frac{1}{\sqrt2} \left( u_{n+1}^m + i v_{n+1}^m \right)\\
v_n^{m+1} &= \frac{1}{\sqrt2} \left( v_{n-1}^m + i u_{n-1}^m \right) \exp{ \left( i n \alpha \right)}\,,
\label{eq:rekursion}
\end{split}
\end{equation}
where $\alpha$ is the induced phase shift between two positions $n$. In addition, a phase shift of $\frac{\pi}{2}$ acquired in the coupler has been taken into account. The discrete time QW of single photons is governed by the same equations \cite{Knight2003, Kempe2003}. Moreover, the dynamics in our system are very similar to classical photon evolution in WGAs \cite{Christodoulides2003, Strauch2006a, Childs2009c}. 

Without phase modulation ($\alpha = 0$), we observe a ballistic spreading of the light field (Figs. 2a-d) as it is well-known from continuous WGAs \cite{Pertsch2002} and QWs \cite{Kempe2003}. The initial asymmetry which is caused by injecting the pulse only into the lower loop remains conserved (Figs. 2c,d). 

Applying a phase shift ($\alpha \neq 0$) which grows linearly in position to the $\ket{\downarrow}$ state at every propagation step $m$ leads to discrete-time photonic Bloch oscillations \cite{Wojcik2004}. Similar to WGAs \cite{Morandotti1999,Pertsch1999}, the field distribution recovers in a quasi-periodic fashion, as it is nicely demonstrated by our experiments (Figs. 2 e,g) and confirmed by simulations (Figs. 2 f,h). 
\begin{figure*}
\includegraphics{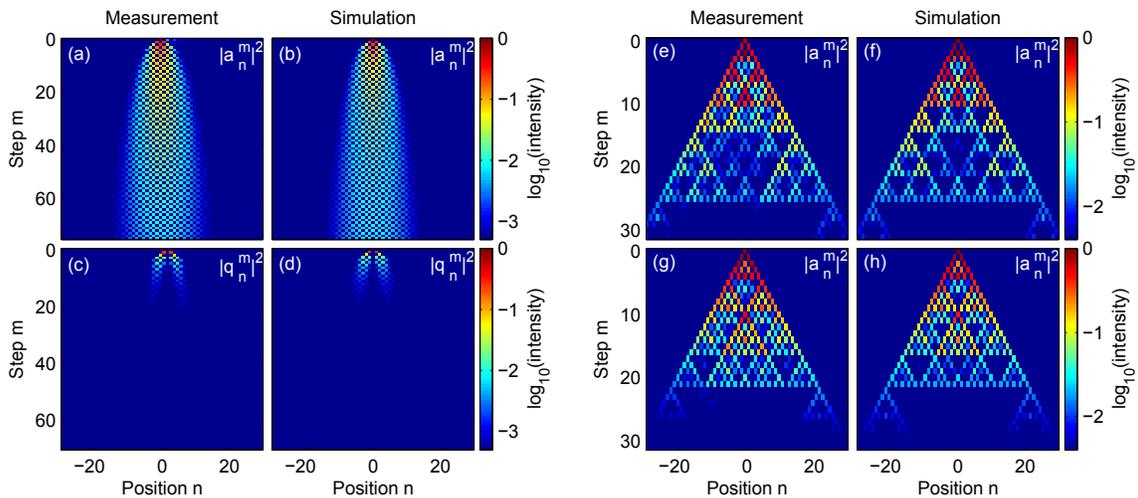}
\caption{\label{fig:lossy} Evolution of wave starting in state $\ket{\downarrow}$ at position $n=0$ in the case with losses (Fig. 1b); (a,b) measured and simulated intensities $\left| a_n^m \right|^2$ in the loop with no phase modulation and (c,d) intensities $\left| q_n^m \right|^2$ at the output port;  (e,f) measured and simulated fractal patterns of intensity $\left| a_n^m \right|^2$ for linear phase gradient $\alpha = \frac{7 \pi}{26}$, net gain $g \approx 3.4$/step; (g,h) $\alpha = \frac{9 \pi}{22}$, $g \approx 3.3$ }
\end{figure*}

So far, on the large scale our lossless system shows the same effects as continuous-time WGAs which is also expected from the fact that discrete-time and continuous-time QWs have the same limiting distribution \cite{Strauch2006a, Childs2009c}.

The situation changes completely if we introduce controlled losses by periodically removing all intensity in the $\ket{\uparrow}$ state. We realize this by a destructive measurement of the $\ket{\uparrow}$ intensity every second time the light has passed the central 50/50 coupler. In a slight modification of our measurement setup (Fig. 1b), we add a second 50/50 coupler and connect the two couplers with fiber pieces that have a length difference $2 \Delta L$, thus realizing the step in position space. Only one of the output ports of the additional coupler is fed back into the loop, whereas all intensity in the second output port is directed to a photodiode. Here, a step $m$ corresponds to a full roundtrip of the light in the loop, i.e. to passing both 50/50 couplers and the light propagation is governed by the equations 
\begin{equation}
\begin{split}
a_n^{m+1} &= \frac{i}{2} \left( a_{n-1}^m + a_{n+1}^m \right) \exp{ \left( i n \alpha \right)}\\
q_n^{m+1} &= \frac{1}{2} \left( a_{n-1}^m - a_{n+1}^m \right)\,,
\label{eq:lossrekursion}
\end{split}
\end{equation}
where $a_n^m$ is the light amplitude inside the loop and $q_n^m$ is the amplitude that is dissipated at the output port at position $n$ and step $m$. Again, we assume that all experimental losses stemming from absorption or monitoring are exactly compensated by the amplifiers. However, a compensation of the photons that are intentionally dissipated at the output port is not possible, because whole pulses leave the system and all information which they carry (amplitude and phase) is completely lost (see \cite{suppl2}). Due to its transverse degrees of freedom, the loss induced evolution in our system is by far more complex than that observed in e.g. two coupled harmonic lossy oscillators \cite{Weinreich1977, Abdullaev2011}.
 
In the absence of phase modulation ($\alpha = 0$), equation (2) is similar to a discretized diffusion equation for the amplitudes $a_n^m$. Moreover, the output amplitudes $q_n^{m+1}$ are the discrete derivative of $a_n^m$. For large iterations ($m \gg 1$), the distribution of amplitudes converges to a Gaussian shape $a_n^m \approx \frac{i^m}{\sqrt{ \frac{\pi}{2} m}} \exp{\left( - \frac{n^2}{2m} \right)}$. The ballistic spreading in the lossless case is thus transformed into a diffusive one. Our measurements (Fig. 3a) demonstrate a linear increase of the variance (Fig. 4a) which confirms the diffusive evolution. At the output port we observe the distribution's discrete derivative (Fig. 3c). Again, the agreement with simulations is very good (Figs. 3b,d). 

\begin{figure}
\includegraphics[width=\columnwidth]{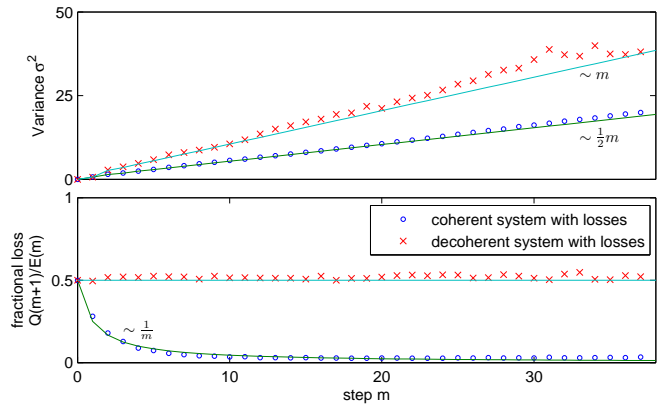}
\caption{\label{fig:decoherence} Transition to a classical random walk by adding decoherence in the form of random phase shifts; (a) variance of Gaussian fit to measured intensity distribution in the coherent lossy system (Fig. 1b) (blue circles) and to the measured average distribution in the case of decoherence (red crosses); lines: simulation of the coherent system and of a classical random walk; (b) fractional energy loss per step to visualize the low losses of the coherent system and the loss of $1/2$ per propagation step $m$ in the case of decoherence.}
\end{figure}
\begin{figure*}
\includegraphics{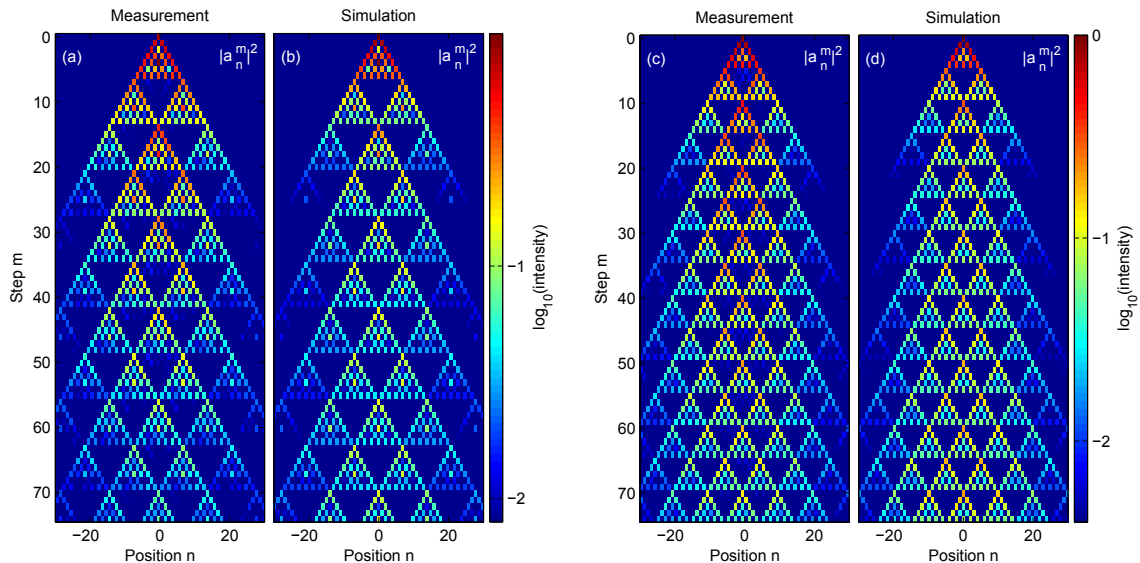}
\caption{\label{fig:fractals} Lossy system with linearly increasing phase shift $\alpha$; (a, b) measured and simulated intensities $\left| a_n^m \right|^2$ in the loop for $\alpha = \frac{3 \pi}{7}$; net gain $g \approx 3.3$ per step; (c, d) $\alpha = \frac{2 \pi}{5}$; $g \approx 3.1$ }
\end{figure*}
One remarkable feature of this system is a surprisingly low rate of photon losses. Looking at the setup in Fig. 1b, one could expect that at each roundtrip half of the photons are coupled out to the photodiode, corresponding to a loss of $1/2$ per propagation step. But owing to wave-mechanical interference, the outcome is completely different. Assuming the absence of all losses except for the selectively introduced ones at the output port, the energy $E(m) \sim \sum\limits_n{\left| a_n^m \right|^2}$ that remains inside the loop at step $m$ decays as slow as $E(m) \sim \frac{1}{\sqrt{\pi m}}$ as is found by integrating the Gaussian shape. This means that $E(m)$ decays slower than any exponential function despite the open port of the second 50/50 coupler in Fig. 1b. The energy $Q(m) \sim \sum\limits_n{\left| q_n^m \right|^2} \sim \frac{1}{\sqrt{m^3}}$ that is dissipated at the output port at step $m$ decreases much faster than $E(m)$, leading to a fractional loss of $\frac{Q(m+1)}{E(m)} \sim \frac1m$ which is quantitatively confirmed by our measurements (Fig. 4b). A system with periodic boundary conditions in position space even possesses a completely lossless mode (see \cite{suppl2}). It is important to note that the photon losses in our system do not lead to decoherence \cite{KENDON2007,Gonulol2009,Schreiber2011a,Broome2010}, but instead a high degree of ordering can be observed.  The astonishing stability of the field in the loop is quite analog to the quantum mechanical Zeno effect \cite{Abdullaev2011}, where a permanent measurement prevents a state from decaying.

If we intentionally remove the coherence from our system by adding random phases \cite{KENDON2007,Schreiber2011a} to the light pulses at every step and position, a photon gets lost at the output port with a probability of 1/2 per step as expected classically (Fig. 4b). The average dynamics approach a classical random walk, where the intensities $\left| a_n^m \right|^2$ perform the diffusion process instead of the amplitudes $a_n^m$. This acceleration due to phase noise \cite{Caruso2009} is also supported by a roughly twofold increase of the variance (Fig. 4a). 

The introduction of a phase gradient $\alpha = \frac{p}{q} \pi$, with coprime $p$ and $q$, also leads to an exponential decay of $E(m)$. In contrast to the lossless system (Figs. 2e-h), no Bloch oscillations are observed. Instead, we discover a completely new and unexpected behavior in the case of losses: The intensity $\left| a_n^m \right|^2$ evolves in a triangular pattern. The field $a_n^m$ below the basis of each triangle of length $q$ is exactly zero because of symmetries in the interference of all possible photon paths (see \cite{suppl2}). With increasing $p$ and $q$, the intensity pattern inside the triangles acquires a complex fractal nature with similarities to the patterns known from linear cellular automata \cite{Wolfram1983,TAKAHASHI1992,Willson1984}. An analytical approach to further describe the observed patterns will be discussed elsewhere. Our measurements shown in Figs. 3e,g and 5a,c are, to our knowledge, the first experimental demonstration of this type of fractal pattern in optics. The rich structure of the arising fractals is confirmed by simulations (Figs. 3f,h and 5b,d).

In summary, we demonstrated an unexpected role of photon losses in an optical system that is discrete in both time and space, linking our observation to photonic waveguide arrays and quantum walks. The introduction of state-selective losses leads to the new phenomena of sub-exponential losses and fractal evolution patterns which have no direct counterpart in continuous WGAs or QWs. All discussed effects have been demonstrated experimentally and confirmed by simulations. Our work gives a new perspective to particle losses in classical and quantum systems. This will trigger a thorough search for other types of such new effects in linear and nonlinear optics, possibly with the help of methods known from cellular automaton research. Apart from this, the experimentally realized fiber loop schemes might likewise be applied to control multi-pulse regimes in fiber ring lasers.

\begin{acknowledgments}
We acknowledge fruitful discussions with A. T. Baraviera, C. Metzner and F. Marquardt. Moreover, we acknowledge financial support from DFG Forschergruppe 760, the German-Israeli Foundation and the Cluster of Excellence Engineering of Advanced Materials (EAM).
\end{acknowledgments}

\end{document}